\documentclass[fleqn,usenatbib]{mnras}

\usepackage{newtxtext,newtxmath}
\usepackage[T1]{fontenc}
\usepackage{ae,aecompl}

\usepackage{graphicx}	
\usepackage{amsmath}
\usepackage{amssymb}

\title[Exploring the DM of the Milky Way]{Exploring the dispersion measure of the Milky Way halo}

\author[Keating \& Pen]{Laura C. Keating$^1$\thanks{lkeating@cita.utoronto.ca} and Ue-Li Pen$^{1,2,3,4,5}$
\\
$^{1}$Canadian Institute for Theoretical Astrophysics, University of Toronto, 60 St. George Street, Toronto, ON M5S 3H8, Canada\\
$^{2}$Canadian Institute for Advanced Research, CIFAR Program in Gravitation and Cosmology, Toronto, ON M5G 1Z8, Canada\\
$^{3}$Dunlap Institute for Astronomy \& Astrophysics, University of Toronto,
AB 120-50 St. George Street, Toronto, ON M5S 3H4,\\
Canada\\
$^{4}$Perimeter Institute of Theoretical Physics, 31 Caroline Street North, Waterloo, ON N2L 2Y5, Canada\\
$^{5}$Max-Planck-Institut f\"ur Radioastronomie, Auf dem H\"ugel 69, D-53121 Bonn, Germany}

\date{Accepted XXX. Received YYY; in original form ZZZ}

\pubyear{xxx}

\begin{document}
\label{firstpage}
\pagerange{\pageref{firstpage}--\pageref{lastpage}}
\maketitle

\begin{abstract}
Fast radio bursts offer the opportunity to place new constraints on the mass and density profile of hot and ionized gas in galactic haloes. We test here the X-ray emission and dispersion measure predicted by different gas profiles for the halo of the Milky Way. We examine a range of models, including entropy stability conditions and external pressure continuity.  We find that incorporating constraints from X-ray observations leads to favouring dispersion measures on the lower end of the range given by these models.  We show that the dispersion measure of the Milky Way halo could be less than 10 cm$^{-3}$ pc in the most extreme model we consider, which is based on constraints from \ion{O}{vii} absorption lines. However, the models allowed by the soft X-ray constraints span more than an order of magnitude in dispersion measures. Additional information on the distribution of gas in the Milky Way halo could be obtained from the signature of a dipole in the dispersion measure of fast radio bursts across the sky, but this will be a small effect for most cases.
\end{abstract}



\section{Introduction}
Over the last decade, the number of known fast radio bursts (FRBs) has grown significantly \citep[see][for a recent review]{petroff2019}, with almost 1000 discovered by the CHIME telscope alone \citep{fonseca2020}. FRBs have a short duration (up to a few milliseconds), are bright (up to a few hundred Jy) and have been detected over a wide frequency range (400 MHz--400 GHz). The origin of these FRBs still remains to be confirmed, but much can be learnt from their observed signal. As FRBs propagate through ionized gas, they interact with free electrons which cause the signal to disperse; that is, a frequency-dependent shift in the group velocity of the wave is observed. The dispersion measure (DM) of a FRB is therefore a probe of the integrated electron density along the line of sight. FRBs have been confirmed to have an extragalactic origin, due to the identification of host galaxies \citep[e.g.,][]{tendulkar2017,mahony2018} as well as statistical arguments based on clustering redshifts \citep{li2019}. Their DMs will therefore have contributions from both the interstellar medium (ISM) and circumgalactic gas of the host galaxy and the Milky Way, as well as intergalactic gas and intervening haloes \citep[see][hereafter PZ19, for a comprehensive review]{prochaska2019}.

It is thought that there may be a significant reservoir of hot baryons in the circumgalactic medium (CGM) of halos which would resolve the ``missing baryon problem''. If these baryons do exist in hot gas ($T \gtrsim 10^6$ K) in the CGM of galaxies, they should be detectable in their soft X-ray emission \citep{anderson2010,anderson2013}. In our Galaxy, there are constraints from the homogeneous component of the soft X-ray background \citep{hasinger1993, moretti2003} or contributions to X-ray emission from a hot halo component \citep{henley2013}. By modelling this emission, it is possible to constrain the distribution of gas in the CGM of the Milky Way \citep{pen1999,fang2013,yamasaki2020}.

FRBs offer another probe of this hot gas, as if there is a significant reservoir of hot, ionized circumgalactic baryons in galactic halos then this will contribute to the total observed DM \citep{mquinn2014,munoz2018}. Upper limits on the density profile of the Galactic CGM can be obtained from localised FRBs with very low DMs \citep[e.g.,  FRB181030,][]{chimefrb2019}. A larger sample of FRBs at known redshifts could also start to place constraints on allowed gas profiles  \citep{mquinn2014}. As well as the DM, additional constraints from the rotation measure and pulse widths of FRBs are also possible \citep{prochaska2019obs}, and indeed first studies using these methods already hint that the CGM may be more diffuse than suggested by other works \citep{cantalupo2014,lan2017}. 

Even the contribution to an observed FRB DM from the halo of our Galaxy is somewhat uncertain. If the gas in our Galaxy's halo traced the dark matter at the cosmic baryon fraction, the DMs of known, localized FRBs would be much higher than observed. The standard resolution is to invoke the expected feedback processes to reduce the column of gas between the Sun and the edge of the Milky Way halo. One might expect this to result in a broad allowed range of halo DMs that can be very small, and bounded from above by observational constraints, including FRB DMs, X-ray background, hydrostatic equilibrium, and external pressure balance. It is perhaps then surprising that a narrow range for the DM of the Galactic halo has been suggested (e.g., PZ19).  

\begin{figure}
\centering
\includegraphics[width=0.96\columnwidth]{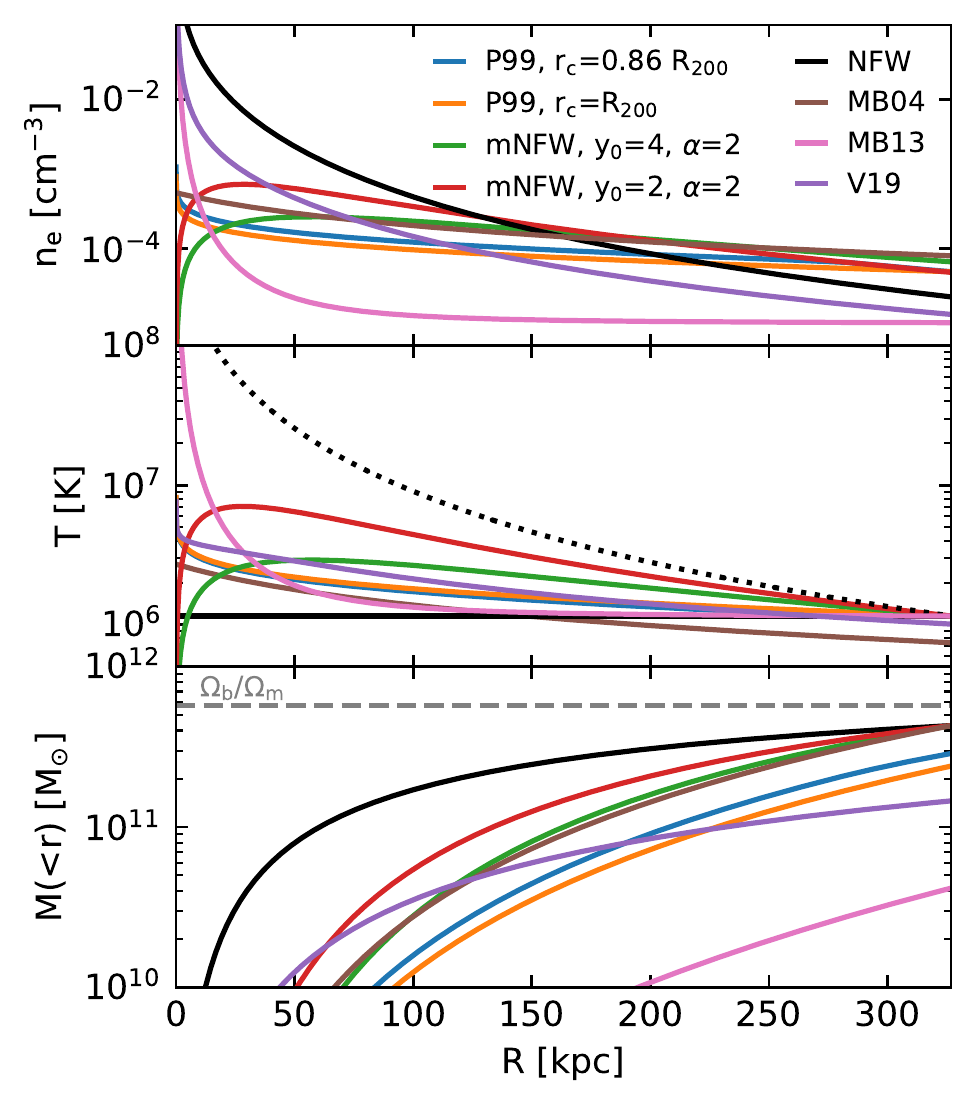}
    \caption{Top: Electron number density as a function of radius for the gas profiles we consider in a halo with $M_{\rm MW halo} = 3.5 \times 10^{12} M_{\odot}$. We show two cases for the \citet{pen1999} model: one where the core radius is chosen to match the upper limit on the unresolved X-ray background as discussed in section \ref{sec:xray} (blue line) and one where the core radius is equal to the virial radius of the halo (orange line). Middle: Temperature as a function of radius. We show two cases for the temperature of the NFW profile: constant entropy (dotted line) and constant temperature (solid line). Bottom: Enclosed gas mass as a function of radius. The horizontal grey dashed line shows the cosmological gas fraction.}
    \label{fig:dens}
\end{figure}

In this Letter, we explore the allowed range of values for the DM due to the ionized gas in the halo of the Milky Way, incorporating constraints from soft X-ray emission. In particular, we revisit some of the halo density profiles presented in PZ19, as well as some additional profiles. We begin by describing these models in section \ref{sec:profiles}. We compute their expected X-ray emission in section  \ref{sec:xray} and their DM in section \ref{sec:dm}. Finally, we present our conclusions in section \ref{sec:conclusions}.

\section{Halo Gas Density Profiles}
\label{sec:profiles} 

It is instructive to first consider simple limits.  The Milky Way infall radius is about 2 Mpc/$h$, which collapses by a factor of 9 into the virial radius of 220 kpc/$h$ (for a halo with a constant rotation velocity of 220 km s$^{-1}$).  The fiducial DM through the halo would be 0.07 cm$^{-3}$ pc if the gas stayed at constant density. For a constant rotation velocity, the mass distribution is isothermal with $\rho \propto 1/r^2$. In ideal adiabatic collapse simulations, gas traces dark matter at or near the cosmic ratio down to ten per cent of the virial radius \citep{frenk1999} and half the cosmic value at small radii. This would result in a DM of over 350 cm$^{-3}$ pc as seen from the solar radius, which is at odds with numerous observations, including FRB DMs and the X-ray background as discussed in \citet{fang2013}. Thus, all models invoke some form of feedback to reduce the amount of gas near the solar radius. One might expect that any value would then be allowed, depending on the assumed feedback process. It seems perhaps surprising that PZ19 find a narrow range of DM. We attribute this to their assumption that gas cannot leave the fiducial halo boundary.  In this case, with infinite heat injection, the gas becomes distributed uniformly within the virial radius, and the DM seen from the centre is 47 cm$^{-3}$ pc: close to the lower value considered by PZ19.

In this Letter, we will open the scenarios to allow gas escape, and impose pressure equilibrium and convective stability.  Perhaps not surprisingly, the allowed range increases substantially. We revisit here several of the halo gas profiles presented in PZ19\footnote{which they have made available on \href{https://github.com/FRBs}{https://github.com/FRBs}}. Several of these models are based on \citet*{nfw1997} (NFW) profile, derived by fitting to density profiles of haloes in dark matter simulations. \citet{mathews2017} presented a modified version of the NFW profile (mNFW), motivated by the impact that baryonic feedback effects would have on the gas density profile and observations of circumgalactic \ion{O}{vi} absorbers. The mNFW profile introduces two additional parameters ($y_0$ and $\alpha$) and is defined as 
\begin{equation}
    \rho(r) = \frac{\rho_0}{y(y_0 + y)^{2 + \alpha}}.
\end{equation}
where $\rho(r)$ is the density at radius $r$, $\rho_0$ is a characteristic density and $y = c\frac{r}{R_{200}}$, where $c$ is the halo concentration and $R_{200}$ is the virial radius. We assume that $c=7.7$ here, following PZ19. For $y_0=1$ and $\alpha=0$ we obtain the original NFW profile. As in PZ19, we consider here mNFW profiles with $y_0=2$ and $y_0=4$, and keep $\alpha=2$ in both cases.

The \citet{maller2004} gas density profile is motivated by the assumption that the halo gas is adiabatic and in hydrostatic equilibrium. Its density profile is defined as 
\begin{equation}
    \rho(r) = \rho_{\rm c} \left( 1 + \frac{3.7}{y} \ln(1 +y) - \frac{3.7}{C_{\rm c}} \ln(1 + C_{\rm c}) \right)^{3/2},
    \end{equation}
where $\rho_{\rm c}$ is a normalisation constant which is determined by the gas mass of the halo, $y$ is defined as above, $C_{\rm c} = c \frac{R_{\rm c}}{R_{200}}$ and $R_{\rm c}$ is the cooling radius, set to 147 kpc as in PZ19.

We also investigate some profiles that were not examined in PZ19. This includes a constraint on the shape of the gas density profile determined from \ion{O}{vii} K$\alpha$ absorption in \citet{miller2013} \citep[see also][for similar constraints using \ion{O}{vii} and \ion{O}{viii} emission lines]{miller2015}. They fit the data with a spherical density model of the form
\begin{equation}
    n(r) = n_0 \left(1 + \left(\frac{r}{r_{\rm c}}\right)^2 \right)^{-3\beta/2},
\end{equation}
with best-fit parameters $n_0$ = 0.46 cm$^{-3}$, $r_{\rm c}$ = 0.35 kpc and $\beta$ = 0.71. They also further add a ambient density component $n_{\rm e}$ = 10$^{-5}$ cm$^{-3}$ at all radii, based on ram-pressure stripping of dwarf spherioidals \citep{blitz2000}.

We also look at the pNFW model of \citet{voit2019}. This model assumes a constant circular velocity at small radii, and the velocity profile of a NFW halo at larger radii and accounts for the precipitation of cold clouds from the hot halo gas. The density is determined from the fitting function 
\begin{equation}
    n(r) = \left\{ \left[ n_1 \left(\frac{r}{1 \rm{kpc}}\right)^{-\zeta_1} \right]^{-2} +  \left[ n_2 \left(\frac{r}{100 \rm{kpc}}\right)^{-\zeta_2} \right]^{-2}  \right\}^{-1/2},
\end{equation}
where $n_1$, $n_2$, $\zeta_1$ and $\zeta_2$ are interpolated from the coefficients tabulated as a function of halo mass in \citet{voit2019}.  

Finally, we consider the entropy-floor singular isothermal sphere model of \citet{pen1999}, which was motivated to match the observed limits on the soft X-ray background and invokes baryonic feedback to create a shallower slope in the inner part of the halo gas density profile. This density profile is defined as 
\begin{equation}
    \rho(r) = \left( \frac{f_{\mathrm{g}} v_{\mathrm{circ}}^2}{4 \pi G} \right)
    \begin{cases}
     \frac{1}{r^{2}} & r > r_{\mathrm{c}}, \\
     \frac{1}{r_{\mathrm{c}}^{2}} \left(1 + \frac{12}{25} \ln \left(\frac{r_{\mathrm{c}}}{r}\right) \right)^{3/2}     & r \leq r_{\mathrm{c}}. \\
    \end{cases}
\end{equation}

Examples of these density profiles are shown in the top panel of  Figure \ref{fig:dens}. Here, we have assumed that the mass of the Milky Way can be calculated from an extrapolation of the circular velocity at the radius of the Sun (as assumed in \citealt{pen1999}). This results in a mass of $M_{\rm MW halo} = 3.5 \times 10^{12} M_{\odot}$, somewhat higher than most estimates of the mass of the Milky Way halo, which at the highest end are about $M_{\rm MW halo} \sim 2.5 \times 10^{12} M_{\odot}$ \citep{li2008,boylankolchin2010}. We use this high halo mass as an upper limit on the gas mass of the Milky Way halo, and also examine a lower mass halo with the properties assumed in PZ19 ($M_{\rm MW halo} = 1.5 \times 10^{12} M_{\odot}$) in sections \ref{sec:xray} and \ref{sec:dm}. We also note that some estimates for the Milky Way halo mass can be as low as $M_{\rm MW halo} \sim 0.5 \times 10^{12} M_{\odot}$ \citep{gibbons2014}, which would lead to even lower gas masses (and hence lower dispersion measures) than the models presented here.

\begin{figure}
\centering
\includegraphics[width=0.94\columnwidth]{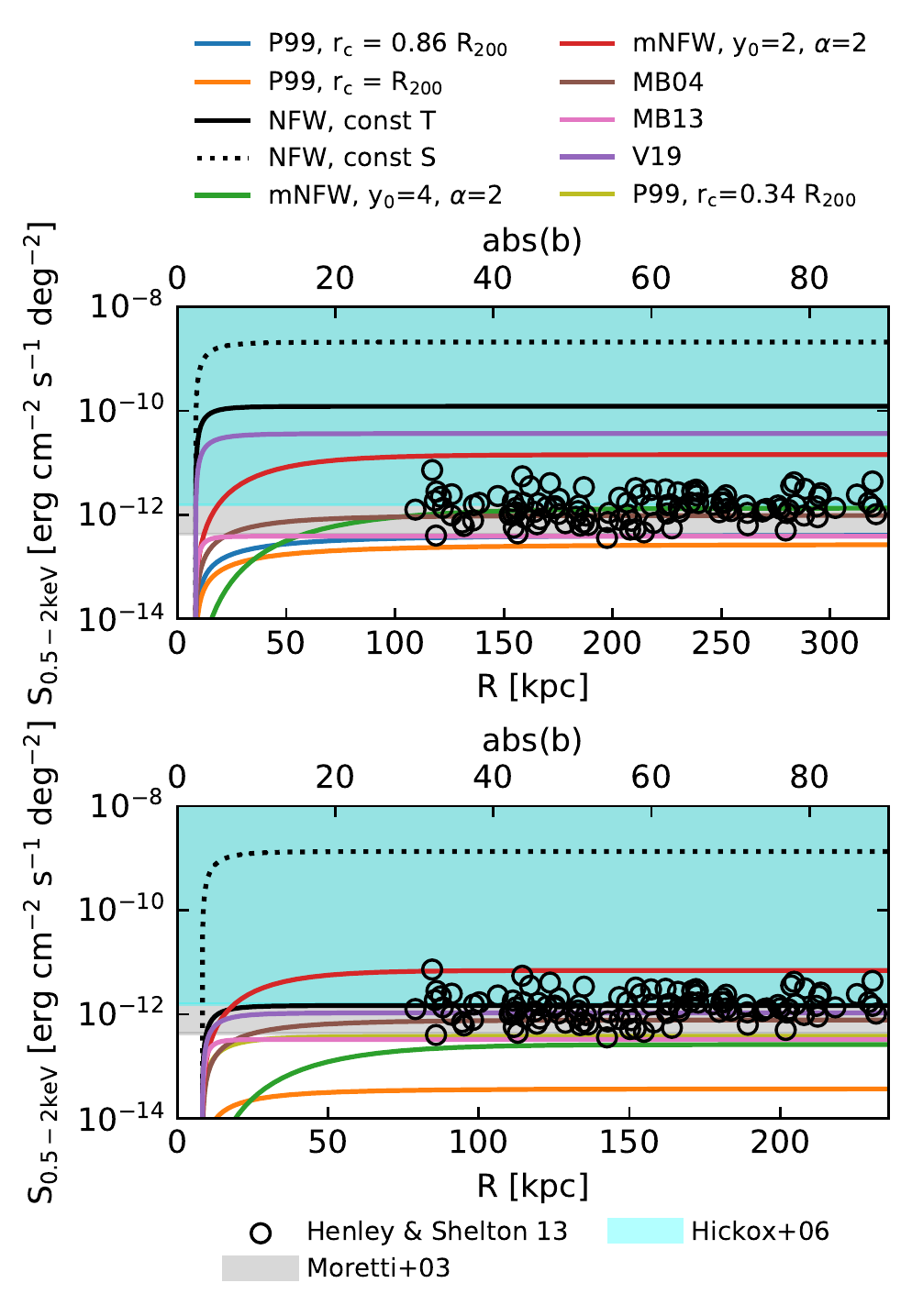}
    \caption{We show the total 0.5--2 keV X-ray emission predicted for each of the gas profiles we consider and for two halo masses (top: $M_{\rm MW halo} = 3.5 \times 10^{12} M_{\odot}$, bottom:  $M_{\rm MW halo} = 1.5 \times 10^{12} M_{\odot}$). The grey and cyan shaded regions show what is excluded by the upper limit on the unresolved X-ray background \citep{moretti2003,hickox2006}. The black circles are estimates of the emission from the Milky Way halo \citep{henley2013} for sightlines plotted at their galactic latitude (top x-axis).}
    \label{fig:xrays}
\end{figure}

In the middle panel of Figure \ref{fig:dens}, we show the temperature as a function of radius we assume for each of these models. For the NFW profile, we take two cases: first, assuming that the halo is isothermal with a temperature given by the virial temperature (black solid line) and second, that the entropy inside the halo is constant (black dashed line). The \citet{pen1999} model assumes that the entropy is constant inside the core radius and that the gas is isothermal and equal to the virial temperature otherwise otherwise. For the \citet{voit2019} model, we calculate the entropy using another fitting function that they provide. We assume constant entropy inside the halo for all other models, with the constraint that the temperature should reach the virial temperature at the virial radius. For the mNFW models, we note that this assumption leads to a non-monotonic temperature profile. 

We show the enclosed gas mass as a function of radius for all models in the bottom panel of Figure \ref{fig:dens}. All of the models presented in PZ19 assume that the halo gas fraction is 0.75 times the cosmological gas fraction. This means that even though feedback is being invoked as a mechanism for reshaping the density profile, it is not actually expelling any gas from the halo. We note that no assumption for a halo gas fraction is needed in \cite{pen1999}, where the halo gas fraction falls naturally as the core radius is increased. The \citet{voit2019} profile similarly has a lower enclosed gas mass. The enclosed mass of the \citet{miller2013} model has by far the lowest enclosed mass, although as noted in that work it is consistent with the expected mass of observed high velocity clouds in the Galactic halo \citep{lehner2012}.

\section{X-ray Emission of the Milky Way Halo}
\label{sec:xray}

We next compute the X-ray emission predicted from each model, and compare it to from constraints from the 0.5--2 keV X-ray band. We assume this emission is dominated by free-free emission, with a volume emissivity 
\begin{equation}
    \epsilon_{\nu}^{\rm ff} = 6.8 \times 10^{-38} Z_i^2 n_{\rm e} n_{i} T^{-1/2} e^{- h_{\rm pl}\nu/k_{\rm B}T}\bar{g} \, \rm{erg} \, \rm{s}^{-1} \, \rm{cm}^{-3} \, \rm{Hz}^{-1},
\end{equation}
as in \citet{rybicki1986}, where $n_e$ and $n_{\rm i}$ are the number densities of electrons and ions, $Z_i$ is the charge of the ion, $T$ is the temperature, $\nu$ is the frequency of interest and $\bar{g}$ is the mean Gaunt factor (which we set to 1.2). $h_{\rm pl}$ and $k_{\rm B}$ are the Planck and Boltzmann constants respectively. We assume that the gas is a mixture of ionized hydrogen and helium. As in \citet{pen1999}, we multiply this by an additional term $c_Z = \frac{Z}{Z_{\odot}} \left(\frac{4 \, \rm{keV}}{T} + 1 \right)$ to account for metal cooling \citep{raymond1976}. Following PZ19, we assume that the halo gas has a metallicity of $Z = 0.3 \, Z_{\odot}$. If the metallicity of halo gas is closer to 0.5  $Z_{\odot}$ \citep{faerman2017,qu2018}, then our predicted X-ray emission would be correspondingly higher, pushing more models closer to the limits of what is allowed by observations. Similarly, if the gas is somewhat hotter than what is assumed in Figure \ref{fig:dens}, then this will also boost the associated X-ray emission. 

We first compute the total X-ray emission in the 0.5--2 keV band, and compare with the estimate of the unresolved X-ray background \citep{moretti2003,hickox2006} as well as  estimates of the soft X-ray emission due to the hot halo of the Milky Way \citep{henley2013} along different sightlines.  For the \cite{pen1999} model, we show cases with a heated core radius that produces X-ray emission at the limit of the observational constraints of \citet{moretti2003}. This corresponds to $r_c = 0.86 \,R_{200}$ for the more massive halo (top panel of Figure \ref{fig:xrays}) and $r_c = 0.34 \, R_{200}$ in the lower mass halo (lower panel). We also show a model with a heated core radius equal to $R_{200}$, the case that maximises the effect of feedback on the CGM of our Galaxy.

We confirm the result of \citet{pen1999} and \citet{fang2013} that both the constant temperature and constant entropy versions of the NFW profile dramatically overproduce the X-ray emission of the Milky Way in all but one of the cases we consider, the isothermal NFW in the lower mass halo. This has a virial temperature $T_{\rm vir} = 6.5 \times 10^{5}$ K, not high enough to produce substantial X-ray emission. Several of the other density profiles we investigate also overproduce the X-ray emission compared to observations, highlighting that this is a useful probe of the CGM. In particular we note that the fiducial halo model of PZ19 (a mNFW profile with $y_0 =2$ and $\alpha=2$) is inconsistent with the X-ray observations, overproducing emission with respect to both the limits from the homogeneous component of the soft X-ray background and the estimates from individual sightlines.

\section{Dispersion Measure of the Milky Way Halo}
\label{sec:dm}

\begin{figure}
\centering
\includegraphics[width=0.94\columnwidth]{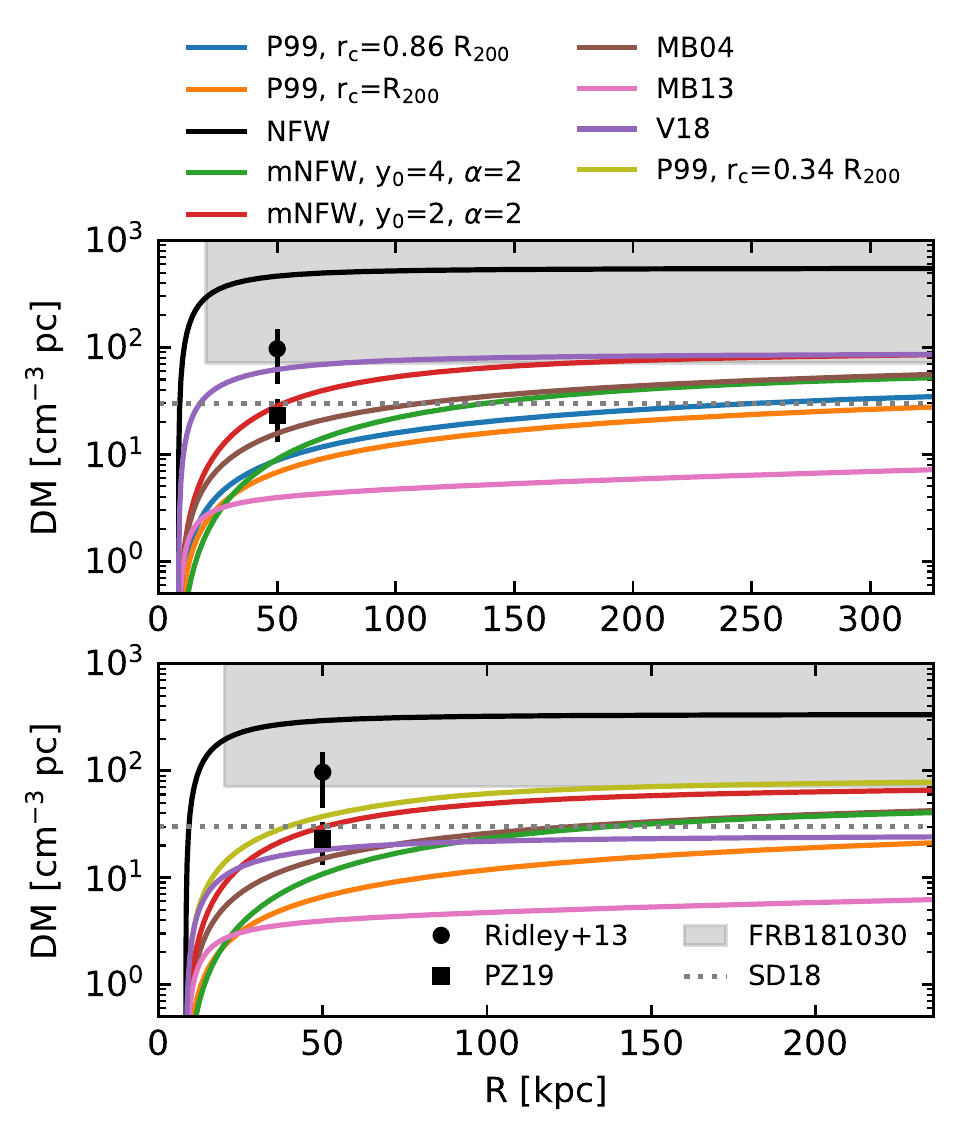}
    \caption{The dispersion measure as a function of radius starting at the radius of the Sun and looking away from the Galactic centre for the two halo masses we consider (top panel: $M_{\rm MW halo} = 3.5 \times 10^{12} M_{\odot}$, bottom panel: $M_{\rm MW halo} = 1.5 \times 10^{12} M_{\odot}$). The horizontal grey dotted line is the DM estimated from \ion{O}{vii} absorption \citep{shull2018}. The circle is the median DM from pulsars in the LMC \citep{ridley2013}, and the errorbars show the complete range of DMs. The square is the halo DM at the distance of the LMC from PZ19. The grey shaded region shows the lowest DM currently measured for a FRB, after the contribution from the Milky Way ISM has been subtracted using the \citet{yao2017} model \citep{chimefrb2019}.}
    \label{fig:dm}
\end{figure}

\begin{figure}
\centering
\includegraphics[width=0.985\columnwidth]{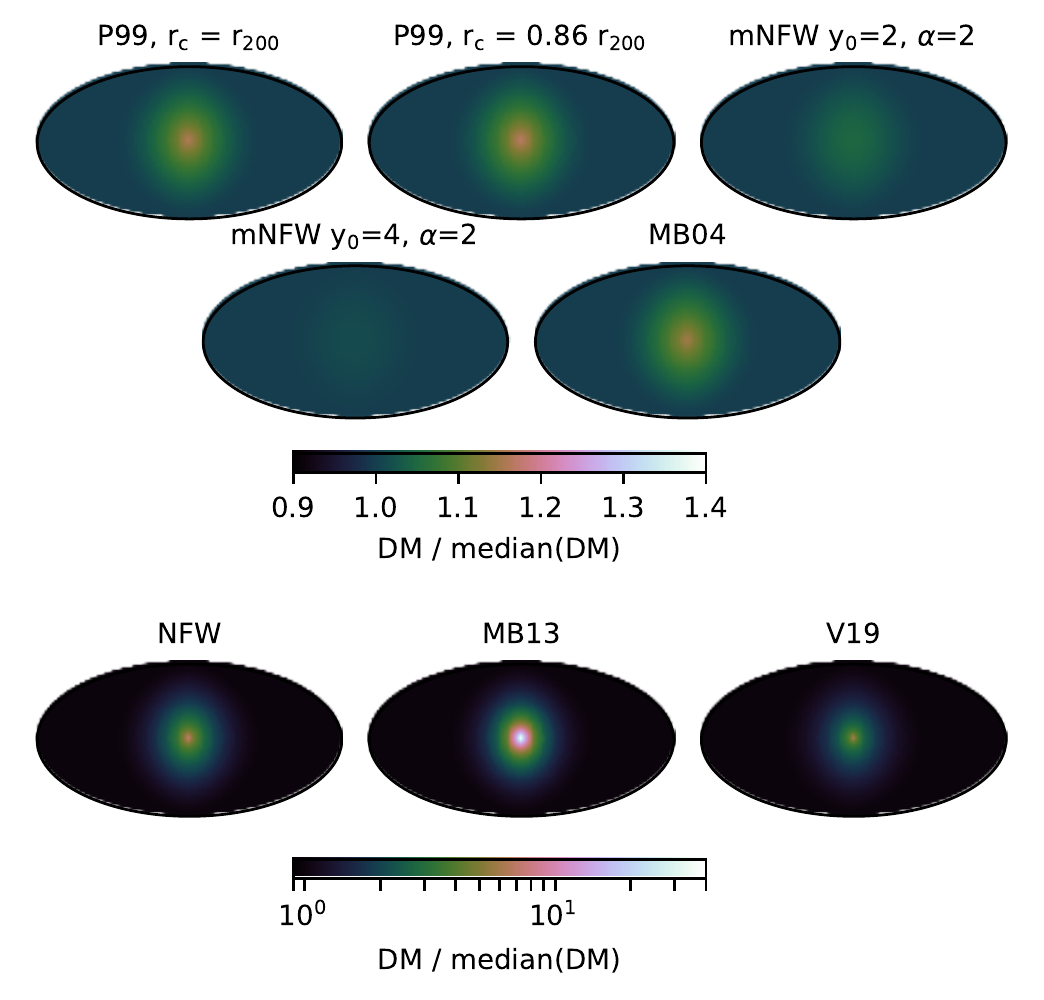}
    \caption{Maps of the change in DM across the sky for the different gas profiles in galactic coordinates for a halo with $M = 3.5 \times 10^{12} \, M_{\odot}$. Each map has been normalised by the median DM predicted for that model. Note that the top two rows all use the same linear colour scale (see top colourbar) and the bottom row uses a logarithmic colour scale (see bottom colourbar). }
    \label{fig:dm_maps}
\end{figure}

We compute the DM predicted for each model, defined as $\rm{DM}  = \int n_{\rm{e}} \, \rm{d}l$, as a function of radius (starting at the solar radius) out to the virial radius of our Galaxy (Figure \ref{fig:dm}). The DMs vary by almost two orders of magnitude depending on the gas density profile assumed, although some of these can already be ruled out based on their X-ray emission (section \ref{sec:xray}). Among the models that are compatible with the observations of \citet{henley2013},  the maximum DM is 55 cm$^{-3}$ pc. The \citet{miller2013} model and the \citet{pen1999} model assuming a heated core radius compatible with the X-ray constraints  predict a DM lower than the estimate of 30 cm$^{-3}$ pc commonly used for the Milky Way halo \citep{dolag2015,shull2018}. This is also the case in the \citet{voit2019} model for the lower mass halo we consider. In the most extreme gas profile we consider (the \citealt{miller2013} profile), we find that the DM could be as low as 6 cm$^{-3}$ pc in the lower mass halo model. This is substantially lower than the DM measured for pulsars in the Large Magellanic Cloud (LMC), but that could be due to the ISM contribution to the DM which is not included here.  There is therefore considerable variation between models, and the DM of the Milky Way halo is perhaps not as well known as claimed in some works. We note that the FRB with the lowest DM currently measured already begins to rule out some models after subtracting the ISM component, even though the quoted DM still contains contributions from intergalactic gas and the FRB host, as well as the Milky Way halo.

We next investigate the dipole distortions of the Milky Way DM, i.e., the fact that it will be higher when looking towards the Galactic centre (Figure \ref{fig:dm_maps}).  We find that, for most of the models, this effect is small with the DM varying by only tens of per cent among different sightlines. For the NFW, \citet{voit2019} and \citet{miller2013} profiles, this effect is more extreme with the DM increasing by a factors of a few in sightlines that are close to the centre of the halo. This effect will likely be difficult to tease out, however, due to the fact that the CGM is likely more complicated than the smooth density profiles we consider here and that in the centre of the halo there will also  be contribution from the ISM, although this has been extensively modelled \citep{cordes2002,cordes2003,gaensler2008,yao2017}.

\section{Conclusions}
\label{sec:conclusions}

We have presented here an investigation into the range of possibilities for the DM of the Milky Way halo. Previous works have claimed that there is only a small uncertainty associated with this quantity. We test a range of models of the gas density profile of the Milky Way CGM against their predicted soft X-ray emission, and find that some are disfavoured by the data. However, even among the allowed models, the predicted values for the halo DM differ by more than an order of magnitude. We note that additional constraints can be obtained from absorption line studies of the Galactic halo and the CGM of other Milky Way-like galaxies, as discussed in detail in PZ19. As more FRBs are localised to host galaxies, they will open an exciting new avenue for probing the CGM of the Milky Way and other galaxies, complementary to absorption and emission line studies.

\section*{Acknowledgements}

We thank Matt McQuinn for helpful comments on a draft of this letter. LCK acknowledges the support of a CITA fellowship. We receive support from Ontario Research Fund-research Excellence Program (ORF-RE), Natural Sciences and Engineering Research Council of Canada (NSERC), Canadian Institute for Advanced Research (CIFAR), Simons Foundation, Thoth Technology Inc, and Alexander von Humboldt Foundation.

\bibliographystyle{mnras}
\bibliography{ref} 

\begin{thebibliography}{}
\makeatletter
\relax
\def\mn@urlcharsother{\let\do\@makeother \do\$\do\&\do\#\do\^\do\_\do\%\do\~}
\def\mn@doi{\begingroup\mn@urlcharsother \@ifnextchar [ {\mn@doi@}
  {\mn@doi@[]}}
\def\mn@doi@[#1]#2{\def\@tempa{#1}\ifx\@tempa\@empty \href
  {http://dx.doi.org/#2} {doi:#2}\else \href {http://dx.doi.org/#2} {#1}\fi
  \endgroup}
\def\mn@eprint#1#2{\mn@eprint@#1:#2::\@nil}
\def\mn@eprint@arXiv#1{\href {http://arxiv.org/abs/#1} {{\tt arXiv:#1}}}
\def\mn@eprint@dblp#1{\href {http://dblp.uni-trier.de/rec/bibtex/#1.xml}
  {dblp:#1}}
\def\mn@eprint@#1:#2:#3:#4\@nil{\def\@tempa {#1}\def\@tempb {#2}\def\@tempc
  {#3}\ifx \@tempc \@empty \let \@tempc \@tempb \let \@tempb \@tempa \fi \ifx
  \@tempb \@empty \def\@tempb {arXiv}\fi \@ifundefined
  {mn@eprint@\@tempb}{\@tempb:\@tempc}{\expandafter \expandafter \csname
  mn@eprint@\@tempb\endcsname \expandafter{\@tempc}}}

\bibitem[\protect\citeauthoryear{{Anderson} \& {Bregman}}{{Anderson} \&
  {Bregman}}{2010}]{anderson2010}
{Anderson} M.~E.,  {Bregman} J.~N.,  2010, \mn@doi [\apj]
  {10.1088/0004-637X/714/1/320}, \href
  {https://ui.adsabs.harvard.edu/abs/2010ApJ...714..320A} {714, 320}

\bibitem[\protect\citeauthoryear{{Anderson}, {Bregman}  \& {Dai}}{{Anderson}
  et~al.}{2013}]{anderson2013}
{Anderson} M.~E.,  {Bregman} J.~N.,   {Dai} X.,  2013, \mn@doi [\apj]
  {10.1088/0004-637X/762/2/106}, \href
  {https://ui.adsabs.harvard.edu/abs/2013ApJ...762..106A} {762, 106}

\bibitem[\protect\citeauthoryear{{Blitz} \& {Robishaw}}{{Blitz} \&
  {Robishaw}}{2000}]{blitz2000}
{Blitz} L.,  {Robishaw} T.,  2000, \mn@doi [\apj] {10.1086/309457}, \href
  {https://ui.adsabs.harvard.edu/abs/2000ApJ...541..675B} {541, 675}

\bibitem[\protect\citeauthoryear{{Boylan-Kolchin}, {Springel}, {White}  \&
  {Jenkins}}{{Boylan-Kolchin} et~al.}{2010}]{boylankolchin2010}
{Boylan-Kolchin} M.,  {Springel} V.,  {White} S. D.~M.,   {Jenkins} A.,  2010,
  \mn@doi [\mnras] {10.1111/j.1365-2966.2010.16774.x}, \href
  {https://ui.adsabs.harvard.edu/abs/2010MNRAS.406..896B} {406, 896}

\bibitem[\protect\citeauthoryear{{CHIME/FRB Collaboration} et~al.,}{{CHIME/FRB
  Collaboration} et~al.}{2019}]{chimefrb2019}
{CHIME/FRB Collaboration} et~al., 2019, \mn@doi [\apjl]
  {10.3847/2041-8213/ab4a80}, \href
  {https://ui.adsabs.harvard.edu/abs/2019ApJ...885L..24C} {885, L24}

\bibitem[\protect\citeauthoryear{{Cantalupo}, {Arrigoni-Battaia}, {Prochaska},
  {Hennawi}  \& {Madau}}{{Cantalupo} et~al.}{2014}]{cantalupo2014}
{Cantalupo} S.,  {Arrigoni-Battaia} F.,  {Prochaska} J.~X.,  {Hennawi} J.~F.,
  {Madau} P.,  2014, \mn@doi [\nat] {10.1038/nature12898}, \href
  {https://ui.adsabs.harvard.edu/abs/2014Natur.506...63C} {506, 63}

\bibitem[\protect\citeauthoryear{{Cordes} \& {Lazio}}{{Cordes} \&
  {Lazio}}{2002}]{cordes2002}
{Cordes} J.~M.,  {Lazio} T.~J.~W.,  2002, arXiv e-prints, \href
  {https://ui.adsabs.harvard.edu/abs/2002astro.ph..7156C} {pp
  astro--ph/0207156}

\bibitem[\protect\citeauthoryear{{Cordes} \& {Lazio}}{{Cordes} \&
  {Lazio}}{2003}]{cordes2003}
{Cordes} J.~M.,  {Lazio} T.~J.~W.,  2003, arXiv e-prints, \href
  {https://ui.adsabs.harvard.edu/abs/2003astro.ph..1598C} {pp
  astro--ph/0301598}

\bibitem[\protect\citeauthoryear{{Dolag}, {Gaensler}, {Beck}  \&
  {Beck}}{{Dolag} et~al.}{2015}]{dolag2015}
{Dolag} K.,  {Gaensler} B.~M.,  {Beck} A.~M.,   {Beck} M.~C.,  2015, \mn@doi
  [\mnras] {10.1093/mnras/stv1190}, \href
  {https://ui.adsabs.harvard.edu/abs/2015MNRAS.451.4277D} {451, 4277}

\bibitem[\protect\citeauthoryear{{Faerman}, {Sternberg}  \& {McKee}}{{Faerman}
  et~al.}{2017}]{faerman2017}
{Faerman} Y.,  {Sternberg} A.,   {McKee} C.~F.,  2017, \mn@doi [\apj]
  {10.3847/1538-4357/835/1/52}, \href
  {https://ui.adsabs.harvard.edu/abs/2017ApJ...835...52F} {835, 52}

\bibitem[\protect\citeauthoryear{{Fang}, {Bullock}  \& {Boylan-Kolchin}}{{Fang}
  et~al.}{2013}]{fang2013}
{Fang} T.,  {Bullock} J.,   {Boylan-Kolchin} M.,  2013, \mn@doi [\apj]
  {10.1088/0004-637X/762/1/20}, \href
  {https://ui.adsabs.harvard.edu/abs/2013ApJ...762...20F} {762, 20}

\bibitem[\protect\citeauthoryear{{Fonseca} et~al.,}{{Fonseca}
  et~al.}{2020}]{fonseca2020}
{Fonseca} E.,  et~al., 2020, \mn@doi [\apjl] {10.3847/2041-8213/ab7208}, \href
  {https://ui.adsabs.harvard.edu/abs/2020ApJ...891L...6F} {891, L6}

\bibitem[\protect\citeauthoryear{{Frenk} et~al.,}{{Frenk}
  et~al.}{1999}]{frenk1999}
{Frenk} C.~S.,  et~al., 1999, \mn@doi [\apj] {10.1086/307908}, \href
  {https://ui.adsabs.harvard.edu/abs/1999ApJ...525..554F} {525, 554}

\bibitem[\protect\citeauthoryear{{Gaensler}, {Madsen}, {Chatterjee}  \&
  {Mao}}{{Gaensler} et~al.}{2008}]{gaensler2008}
{Gaensler} B.~M.,  {Madsen} G.~J.,  {Chatterjee} S.,   {Mao} S.~A.,  2008,
  \mn@doi [\pasa] {10.1071/AS08004}, \href
  {https://ui.adsabs.harvard.edu/abs/2008PASA...25..184G} {25, 184}

\bibitem[\protect\citeauthoryear{{Gibbons}, {Belokurov}  \& {Evans}}{{Gibbons}
  et~al.}{2014}]{gibbons2014}
{Gibbons} S.~L.~J.,  {Belokurov} V.,   {Evans} N.~W.,  2014, \mn@doi [\mnras]
  {10.1093/mnras/stu1986}, \href
  {https://ui.adsabs.harvard.edu/abs/2014MNRAS.445.3788G} {445, 3788}

\bibitem[\protect\citeauthoryear{{Hasinger}, {Burg}, {Giacconi}, {Hartner},
  {Schmidt}, {Trumper}  \& {Zamorani}}{{Hasinger} et~al.}{1993}]{hasinger1993}
{Hasinger} G.,  {Burg} R.,  {Giacconi} R.,  {Hartner} G.,  {Schmidt} M.,
  {Trumper} J.,   {Zamorani} G.,  1993, \aap, \href
  {https://ui.adsabs.harvard.edu/abs/1993A&A...275....1H} {275, 1}

\bibitem[\protect\citeauthoryear{{Henley} \& {Shelton}}{{Henley} \&
  {Shelton}}{2013}]{henley2013}
{Henley} D.~B.,  {Shelton} R.~L.,  2013, \mn@doi [\apj]
  {10.1088/0004-637X/773/2/92}, \href
  {https://ui.adsabs.harvard.edu/abs/2013ApJ...773...92H} {773, 92}

\bibitem[\protect\citeauthoryear{{Hickox} \& {Markevitch}}{{Hickox} \&
  {Markevitch}}{2006}]{hickox2006}
{Hickox} R.~C.,  {Markevitch} M.,  2006, \mn@doi [\apj] {10.1086/504070}, \href
  {https://ui.adsabs.harvard.edu/abs/2006ApJ...645...95H} {645, 95}

\bibitem[\protect\citeauthoryear{{Lan} \& {Fukugita}}{{Lan} \&
  {Fukugita}}{2017}]{lan2017}
{Lan} T.-W.,  {Fukugita} M.,  2017, \mn@doi [\apj] {10.3847/1538-4357/aa93eb},
  \href {https://ui.adsabs.harvard.edu/abs/2017ApJ...850..156L} {850, 156}

\bibitem[\protect\citeauthoryear{{Lehner}, {Howk}, {Thom}, {Fox}, {Tumlinson},
  {Tripp}  \& {Meiring}}{{Lehner} et~al.}{2012}]{lehner2012}
{Lehner} N.,  {Howk} J.~C.,  {Thom} C.,  {Fox} A.~J.,  {Tumlinson} J.,  {Tripp}
  T.~M.,   {Meiring} J.~D.,  2012, \mn@doi [\mnras]
  {10.1111/j.1365-2966.2012.21428.x}, \href
  {https://ui.adsabs.harvard.edu/abs/2012MNRAS.424.2896L} {424, 2896}

\bibitem[\protect\citeauthoryear{{Li} \& {White}}{{Li} \&
  {White}}{2008}]{li2008}
{Li} Y.-S.,  {White} S. D.~M.,  2008, \mn@doi [\mnras]
  {10.1111/j.1365-2966.2007.12748.x}, \href
  {https://ui.adsabs.harvard.edu/abs/2008MNRAS.384.1459L} {384, 1459}

\bibitem[\protect\citeauthoryear{{Li}, {Yalinewich}  \& {Breysse}}{{Li}
  et~al.}{2019}]{li2019}
{Li} D.,  {Yalinewich} A.,   {Breysse} P.~C.,  2019, arXiv e-prints, \href
  {https://ui.adsabs.harvard.edu/abs/2019arXiv190210120L} {p. arXiv:1902.10120}

\bibitem[\protect\citeauthoryear{{Mahony} et~al.,}{{Mahony}
  et~al.}{2018}]{mahony2018}
{Mahony} E.~K.,  et~al., 2018, \mn@doi [\apjl] {10.3847/2041-8213/aae7cb},
  \href {https://ui.adsabs.harvard.edu/abs/2018ApJ...867L..10M} {867, L10}

\bibitem[\protect\citeauthoryear{{Maller} \& {Bullock}}{{Maller} \&
  {Bullock}}{2004}]{maller2004}
{Maller} A.~H.,  {Bullock} J.~S.,  2004, \mn@doi [\mnras]
  {10.1111/j.1365-2966.2004.08349.x}, \href
  {https://ui.adsabs.harvard.edu/abs/2004MNRAS.355..694M} {355, 694}

\bibitem[\protect\citeauthoryear{{Mathews} \& {Prochaska}}{{Mathews} \&
  {Prochaska}}{2017}]{mathews2017}
{Mathews} W.~G.,  {Prochaska} J.~X.,  2017, \mn@doi [\apjl]
  {10.3847/2041-8213/aa8861}, \href
  {https://ui.adsabs.harvard.edu/abs/2017ApJ...846L..24M} {846, L24}

\bibitem[\protect\citeauthoryear{{McQuinn}}{{McQuinn}}{2014}]{mquinn2014}
{McQuinn} M.,  2014, \mn@doi [\apjl] {10.1088/2041-8205/780/2/L33}, \href
  {https://ui.adsabs.harvard.edu/abs/2014ApJ...780L..33M} {780, L33}

\bibitem[\protect\citeauthoryear{{Miller} \& {Bregman}}{{Miller} \&
  {Bregman}}{2013}]{miller2013}
{Miller} M.~J.,  {Bregman} J.~N.,  2013, \mn@doi [\apj]
  {10.1088/0004-637X/770/2/118}, \href
  {https://ui.adsabs.harvard.edu/abs/2013ApJ...770..118M} {770, 118}

\bibitem[\protect\citeauthoryear{{Miller} \& {Bregman}}{{Miller} \&
  {Bregman}}{2015}]{miller2015}
{Miller} M.~J.,  {Bregman} J.~N.,  2015, \mn@doi [\apj]
  {10.1088/0004-637X/800/1/14}, \href
  {https://ui.adsabs.harvard.edu/abs/2015ApJ...800...14M} {800, 14}

\bibitem[\protect\citeauthoryear{{Moretti}, {Campana}, {Lazzati}  \&
  {Tagliaferri}}{{Moretti} et~al.}{2003}]{moretti2003}
{Moretti} A.,  {Campana} S.,  {Lazzati} D.,   {Tagliaferri} G.,  2003, \mn@doi
  [\apj] {10.1086/374335}, \href
  {https://ui.adsabs.harvard.edu/abs/2003ApJ...588..696M} {588, 696}

\bibitem[\protect\citeauthoryear{{Mu{\~n}oz} \& {Loeb}}{{Mu{\~n}oz} \&
  {Loeb}}{2018}]{munoz2018}
{Mu{\~n}oz} J.~B.,  {Loeb} A.,  2018, \mn@doi [\prd]
  {10.1103/PhysRevD.98.103518}, \href
  {https://ui.adsabs.harvard.edu/abs/2018PhRvD..98j3518M} {98, 103518}

\bibitem[\protect\citeauthoryear{{Navarro}, {Frenk}  \& {White}}{{Navarro}
  et~al.}{1997}]{nfw1997}
{Navarro} J.~F.,  {Frenk} C.~S.,   {White} S. D.~M.,  1997, \mn@doi [\apj]
  {10.1086/304888}, \href
  {https://ui.adsabs.harvard.edu/abs/1997ApJ...490..493N} {490, 493}

\bibitem[\protect\citeauthoryear{{Pen}}{{Pen}}{1999}]{pen1999}
{Pen} U.-L.,  1999, \mn@doi [\apjl] {10.1086/311799}, \href
  {https://ui.adsabs.harvard.edu/abs/1999ApJ...510L...1P} {510, L1}

\bibitem[\protect\citeauthoryear{{Petroff}, {Hessels}  \& {Lorimer}}{{Petroff}
  et~al.}{2019}]{petroff2019}
{Petroff} E.,  {Hessels} J.~W.~T.,   {Lorimer} D.~R.,  2019, \mn@doi [\aapr]
  {10.1007/s00159-019-0116-6}, \href
  {https://ui.adsabs.harvard.edu/abs/2019A&ARv..27....4P} {27, 4}

\bibitem[\protect\citeauthoryear{{Prochaska} \& {Zheng}}{{Prochaska} \&
  {Zheng}}{2019}]{prochaska2019}
{Prochaska} J.~X.,  {Zheng} Y.,  2019, \mn@doi [\mnras] {10.1093/mnras/stz261},
  \href {https://ui.adsabs.harvard.edu/abs/2019MNRAS.485..648P} {485, 648}

\bibitem[\protect\citeauthoryear{{Prochaska} et~al.,}{{Prochaska}
  et~al.}{2019}]{prochaska2019obs}
{Prochaska} J.~X.,  et~al., 2019, arXiv e-prints, \href
  {https://ui.adsabs.harvard.edu/abs/2019arXiv190911681P} {p. arXiv:1909.11681}

\bibitem[\protect\citeauthoryear{{Qu} \& {Bregman}}{{Qu} \&
  {Bregman}}{2018}]{qu2018}
{Qu} Z.,  {Bregman} J.~N.,  2018, \mn@doi [\apj] {10.3847/1538-4357/aaafd4},
  \href {https://ui.adsabs.harvard.edu/abs/2018ApJ...856....5Q} {856, 5}

\bibitem[\protect\citeauthoryear{{Raymond}, {Cox}  \& {Smith}}{{Raymond}
  et~al.}{1976}]{raymond1976}
{Raymond} J.~C.,  {Cox} D.~P.,   {Smith} B.~W.,  1976, \mn@doi [\apj]
  {10.1086/154170}, \href
  {https://ui.adsabs.harvard.edu/abs/1976ApJ...204..290R} {204, 290}

\bibitem[\protect\citeauthoryear{{Ridley}, {Crawford}, {Lorimer}, {Bailey},
  {Madden}, {Anella}  \& {Chennamangalam}}{{Ridley} et~al.}{2013}]{ridley2013}
{Ridley} J.~P.,  {Crawford} F.,  {Lorimer} D.~R.,  {Bailey} S.~R.,  {Madden}
  J.~H.,  {Anella} R.,   {Chennamangalam} J.,  2013, \mn@doi [\mnras]
  {10.1093/mnras/stt709}, \href
  {https://ui.adsabs.harvard.edu/abs/2013MNRAS.433..138R} {433, 138}

\bibitem[\protect\citeauthoryear{{Rybicki} \& {Lightman}}{{Rybicki} \&
  {Lightman}}{1986}]{rybicki1986}
{Rybicki} G.~B.,  {Lightman} A.~P.,  1986, {Radiative Processes in
  Astrophysics}.
{John Wiley \& Sons}

\bibitem[\protect\citeauthoryear{{Shull} \& {Danforth}}{{Shull} \&
  {Danforth}}{2018}]{shull2018}
{Shull} J.~M.,  {Danforth} C.~W.,  2018, \mn@doi [\apjl]
  {10.3847/2041-8213/aaa2fa}, \href
  {https://ui.adsabs.harvard.edu/abs/2018ApJ...852L..11S} {852, L11}

\bibitem[\protect\citeauthoryear{{Tendulkar} et~al.,}{{Tendulkar}
  et~al.}{2017}]{tendulkar2017}
{Tendulkar} S.~P.,  et~al., 2017, \mn@doi [\apjl] {10.3847/2041-8213/834/2/L7},
  \href {https://ui.adsabs.harvard.edu/abs/2017ApJ...834L...7T} {834, L7}

\bibitem[\protect\citeauthoryear{{Voit}}{{Voit}}{2019}]{voit2019}
{Voit} G.~M.,  2019, \mn@doi [\apj] {10.3847/1538-4357/ab2bfd}, \href
  {https://ui.adsabs.harvard.edu/abs/2019ApJ...880..139V} {880, 139}

\bibitem[\protect\citeauthoryear{{Yamasaki} \& {Totani}}{{Yamasaki} \&
  {Totani}}{2020}]{yamasaki2020}
{Yamasaki} S.,  {Totani} T.,  2020, \mn@doi [\apj] {10.3847/1538-4357/ab58c4},
  \href {https://ui.adsabs.harvard.edu/abs/2020ApJ...888..105Y} {888, 105}

\bibitem[\protect\citeauthoryear{{Yao}, {Manchester}  \& {Wang}}{{Yao}
  et~al.}{2017}]{yao2017}
{Yao} J.~M.,  {Manchester} R.~N.,   {Wang} N.,  2017, \mn@doi [\apj]
  {10.3847/1538-4357/835/1/29}, \href
  {https://ui.adsabs.harvard.edu/abs/2017ApJ...835...29Y} {835, 29}

\makeatother
\end{thebibliography}
\label{lastpage}
\end{document}